\documentclass[aps,prl,twocolumn,twoside,preprintnumbers,superscriptaddress,floatfix]{revtex4-2}

\usepackage{epsfig}
\usepackage{amsmath}
\usepackage{amssymb}
\usepackage{latexsym}
\usepackage{xspace}
\usepackage[colorlinks=true,linktocpage=true,linkcolor=blue,citecolor=blue,allcolors=blue]{hyperref}
\usepackage{url}
\usepackage[utf8]{inputenc}
\usepackage{indentfirst}
\usepackage{enumerate}
\usepackage{color}
\usepackage{relsize}
\usepackage{graphicx}
\usepackage{dcolumn}
\usepackage{bm}
\usepackage[caption=false,position=top]{subfig}
\usepackage[english]{babel}
\usepackage{url}
\usepackage[colorlinks=true,linktocpage=true,linkcolor=blue,citecolor=blue,allcolors=blue]{hyperref}

\newcommand{\be}{\begin{equation}}
\newcommand{\ee}{\end{equation}}

\begin{document}

\title{Flavor Violating Axions in the Early Universe}

\author{Francesco D'Eramo}
 \email{francesco.deramo@pd.infn.it}
 \author{Seokhoon Yun}
 \email{seokhoon.yun@pd.infn.it}

\affiliation{Dipartimento di Fisica e Astronomia, Universit\`a degli Studi di Padova, Via Marzolo 8, 35131 Padova, Italy}
\affiliation{INFN, Sezione di Padova, Via Marzolo 8, 35131 Padova, Italy}

\date{November 23, 2021}

\begin{abstract}

Flavor violating axion couplings can be in action before recombination, and they can fill the early universe with an additional radiation component. Working within a model-independent framework, we consider an effective field theory for the axion field and quantify axion production. Current cosmological data exclude already a fraction of the available parameter space, and the bounds will improve significantly with future CMB-S4 surveys. Remarkably, we find that future cosmological bounds will be comparable or even stronger than the ones obtained in our terrestrial laboratories.

\end{abstract}

\maketitle



\noindent {\bf Introduction.} Light and weakly-coupled particles arise naturally within motivated theories for physics beyond the standard model (SM). A notable example is the QCD axion~\cite{Peccei:1977np,Peccei:1977hh,Wilczek:1977pj,Weinberg:1977ma}. Peccei-Quinn (PQ) theories provide a dynamical solution to the strong CP problem and a viable dark matter candidate~\cite{Preskill:1982cy,Abbott:1982af,Dine:1982ah}. Typically, the axion field couples to several SM particles~\cite{Kim:2008hd,DiLuzio:2020wdo}, and experimentalists search for the effects of these interactions both in the early universe~\cite{Marsh:2015xka} and in our laboratories~\cite{Graham:2015ouw,Irastorza:2018dyq,Sikivie:2020zpn}.

In this work, we study the cosmological implications of axion couplings that are not diagonal in the SM fermion generations. They can arise from radiative corrections with both loop and GIM suppressions even if the underlying theory preserves flavor~\cite{Choi:2017gpf,Chala:2020wvs,Bauer:2020jbp,Choi:2021kuy,Bonilla:2021ufe}. Alternatively, flavor violation (FV) can be present  at tree-level~\cite{Wilczek:1982rv,Ema:2016ops,Calibbi:2016hwq,Celis:2014iua,Bjorkeroth:2017tsz,Bjorkeroth:2018dzu,Linster:2018avp,Carone:2019lfc,Calibbi:2020jvd}. We remain agnostic about the high-energy origin, and we work within the effective field theory framework
\be
\mathcal{L}^{(a)}_{{\rm FV}} = \frac{\partial_\mu a}{2 f_a} \sum_{\psi_i \neq \psi_j} \, \bar{\psi}_i \gamma^\mu \left( c^V_{\psi_i \psi_j} + c^A_{\psi_i \psi_j} \gamma^5 \right) \psi_j \ .
\label{eq:LaFV}
\ee
The axion field $a$ interacts with up ($\psi_i = u, c, t$) and down ($\psi_i = d, s, b$) quarks, and charged leptons ($\psi_i = e, \mu, \tau$)~\footnote{We neglect neutrinos since the correspondent rates are suppressed by their tiny masses.}. We define the dimensionful parameters
\be
F^{\alpha}_{\psi_i \psi_j} \equiv \frac{2 f_a}{c^\alpha_{\psi_i \psi_ j}} \ , \qquad \qquad \alpha = \{V,A\} \ .
\label{eq:NormalFV}
\ee
The hermiticity condition on $\mathcal{L}^{(a)}_{{\rm FV}}$ implies $F^{\alpha \, *}_{\psi_j \psi_i} = F^{\alpha}_{\psi_i \psi_j}$. 

FV interactions mediate axion production in the early universe, and this cosmic axion background that would manifest itself today in the cosmic microwave background (CMB) anisotropy spectrum is one of the cosmological consequences of the PQ mechanism~\cite{Turner:1986tb,Masso:2002np,Graf:2010tv,Salvio:2013iaa,Dror:2021nyr}. This effect is historically parameterized in terms of an effective number of additional neutrino species $\Delta N_{\rm eff}$, and the Planck collaboration places the best current constraint, $N_{\rm eff} = 2.99 \pm 0.17$~\cite{Aghanim:2018eyx}. Future CMB-S4 surveys will achieve the remarkable sensitivity $\left. \Delta N_{\rm eff}\right|_{1 \sigma  \left(2\sigma\right)} \simeq 0.03  \left(0.06\right)$~\cite{CMB-S4:2016ple,Abazajian:2019eic} that corresponds approximately to the contribution from axions once in thermal equilibrium that decoupled before the electroweak phase transition~\cite{Brust:2013xpv}. These astonishing forecasts motivated recently a noteworthy effort to provide quantitative and reliable predictions~\cite{Baumann:2016wac,Ferreira:2018vjj,DEramo:2018vss,Arias-Aragon:2020qtn,Arias-Aragon:2020shv,Ferreira:2020bpb,Giare:2020vzo,DEramo:2021psx,DEramo:2021lgb,Giare:2021cqr,Green:2021hjh}.

\begin{figure}
\centering
\includegraphics[width=0.235\textwidth]{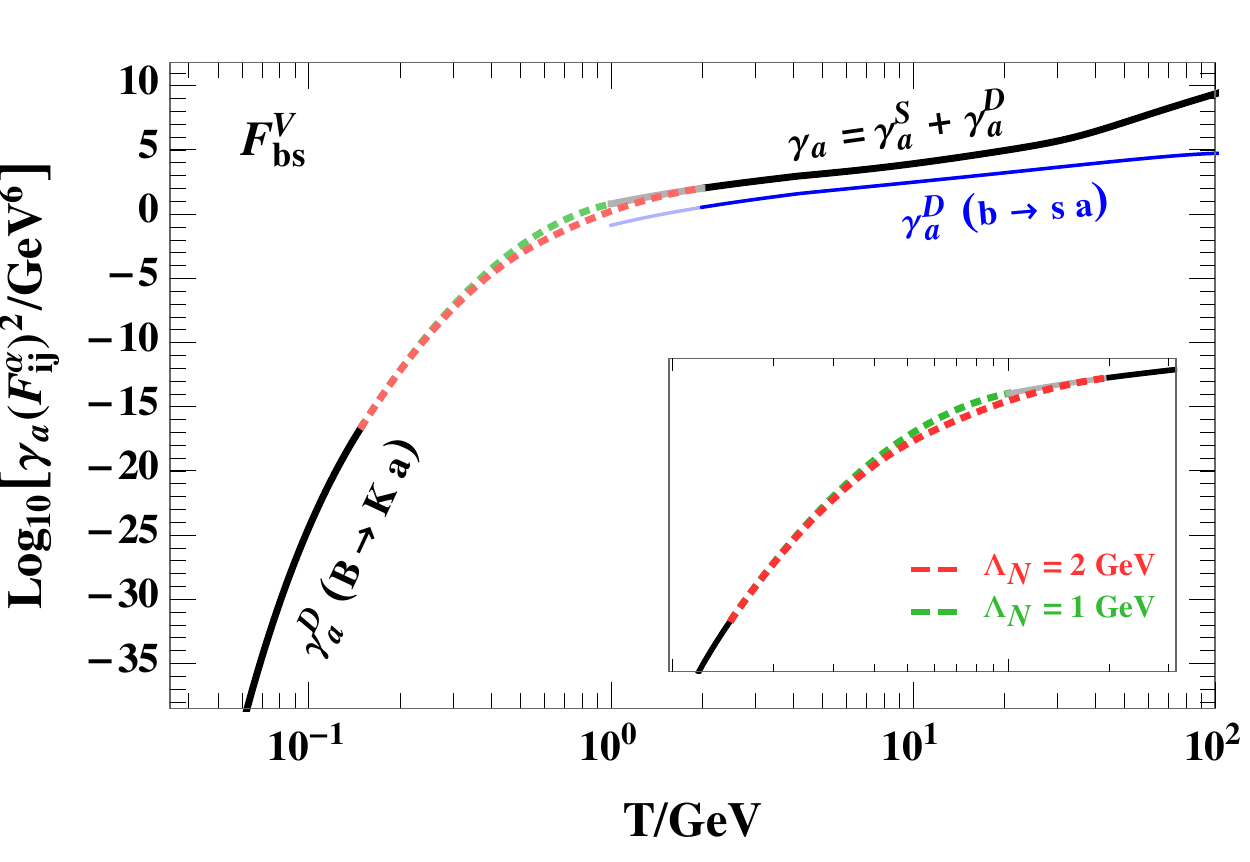}
\includegraphics[width=0.235\textwidth]{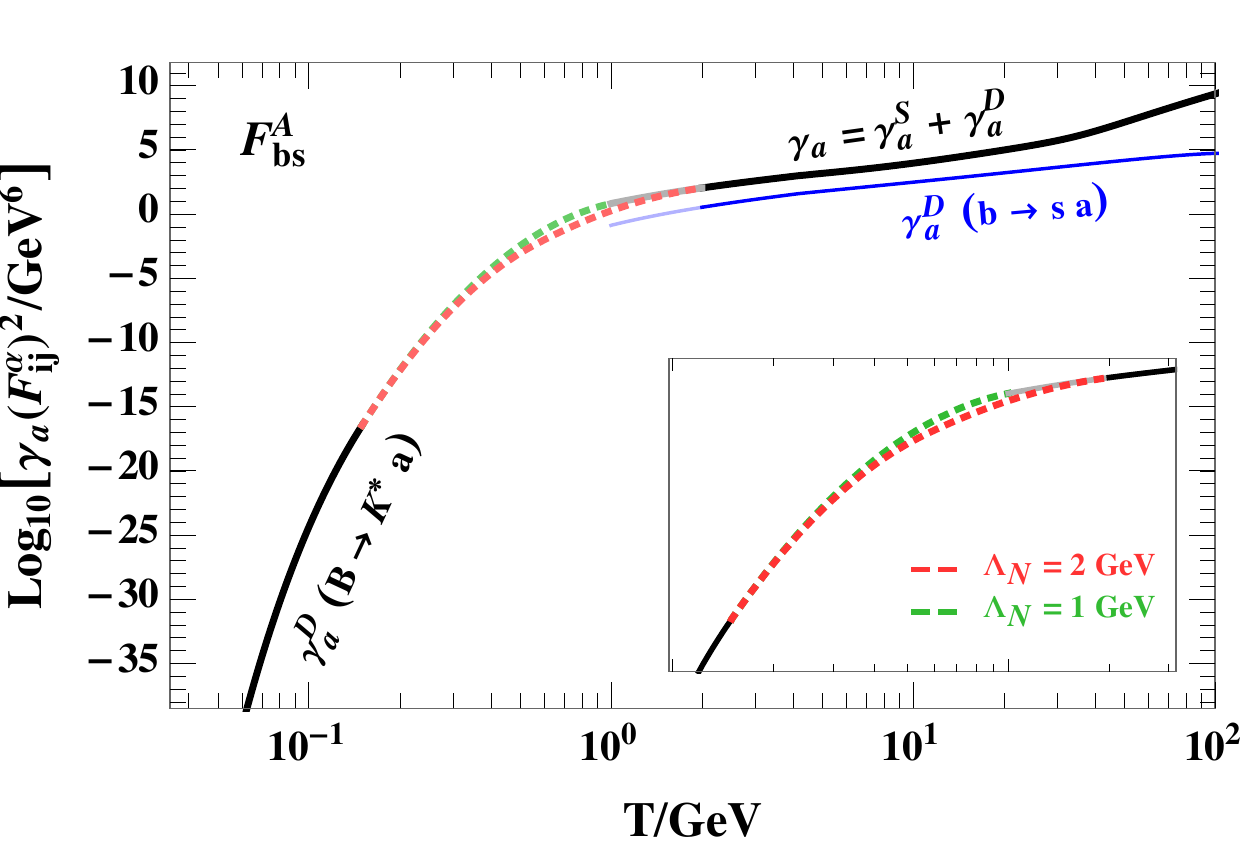}
\includegraphics[width=0.235\textwidth]{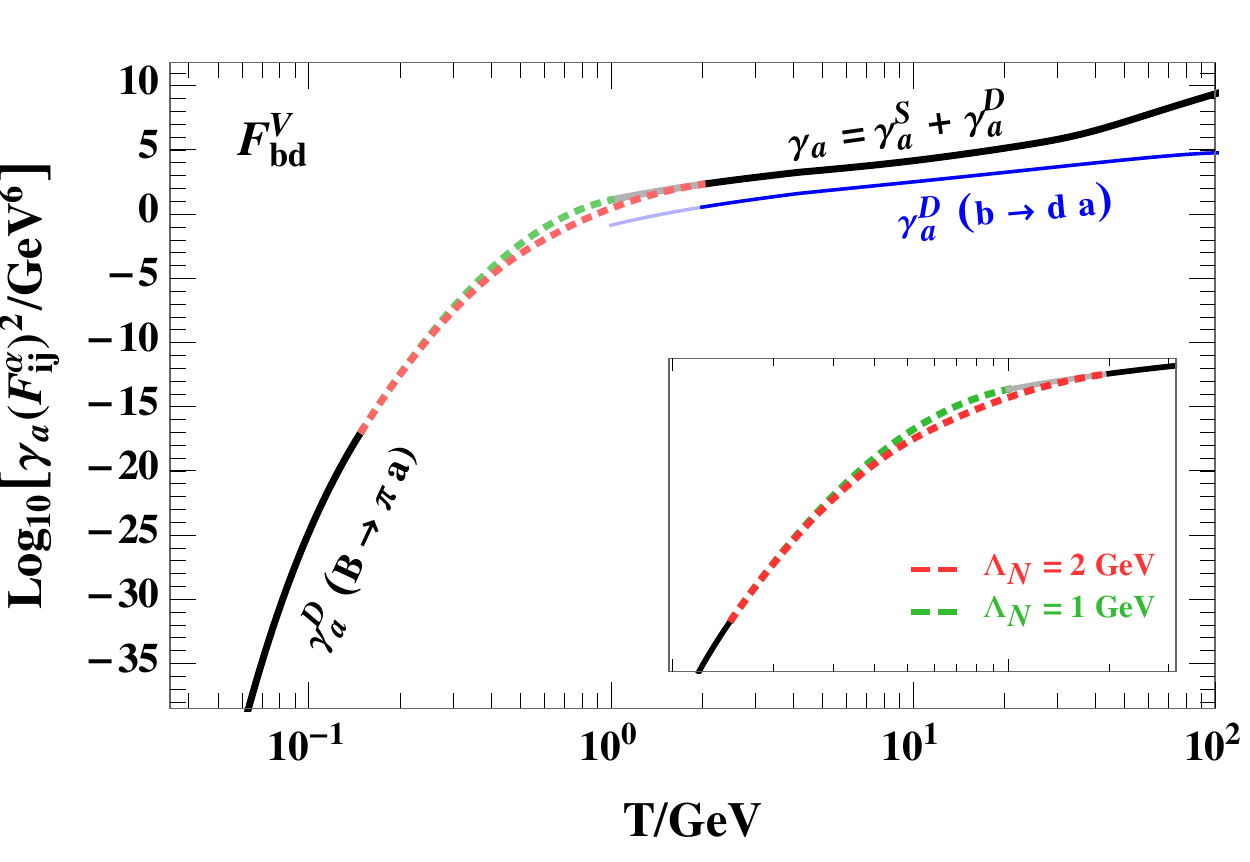}
\includegraphics[width=0.235\textwidth]{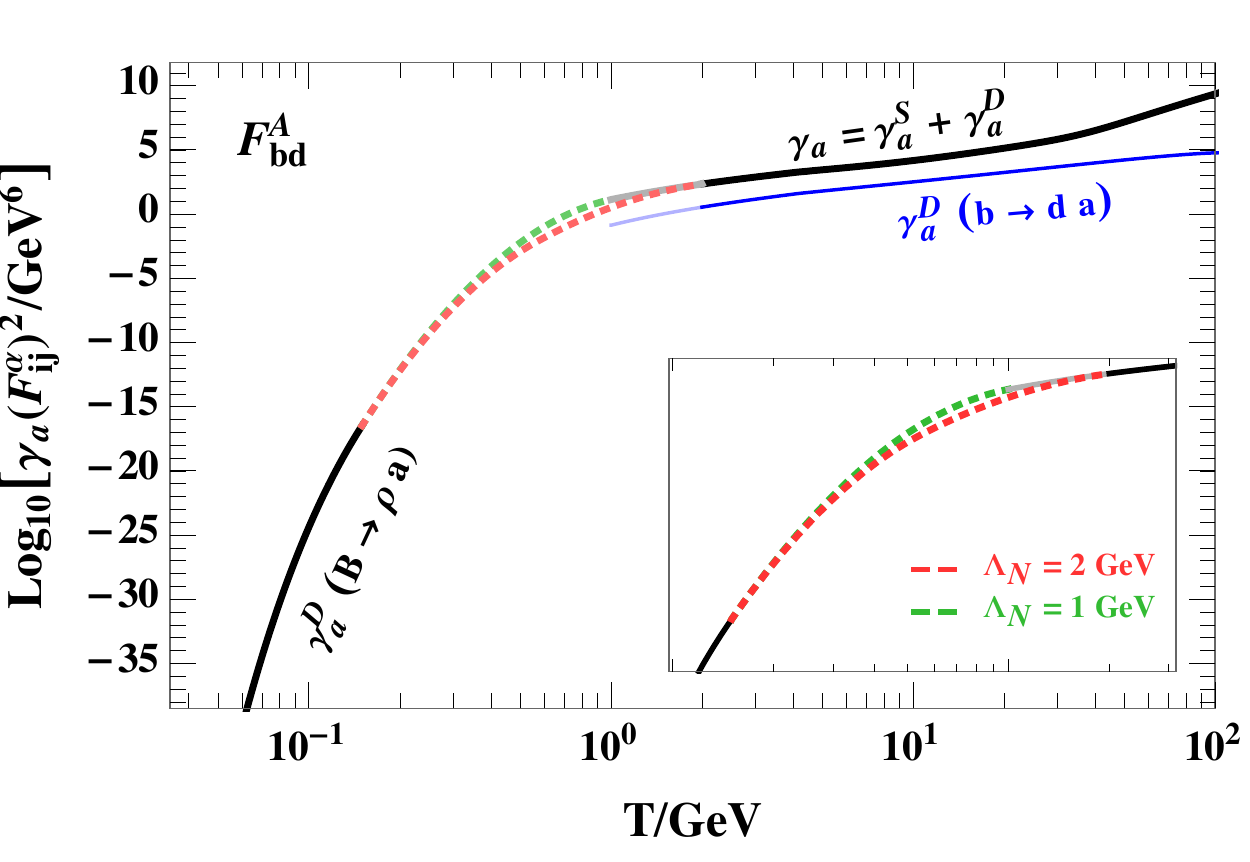}
\caption{\em Axion production rate via FV couplings to $b$ quarks. Black and blue lines denote total and decay rates, respectively. Solid lines stop at the perturbative scale $\Lambda_N = (1, 2 ) \,{\rm GeV}$. The rate between $\Lambda_{\rm HRG} = 150\,{\rm MeV}$ and $\Lambda_N$ is connected smoothly via the cubic spline interpolation (dashed lines).}
\label{fig:Rate3rdQuarks}
\end{figure}

\begin{figure}
\centering
\includegraphics[width=0.235\textwidth]{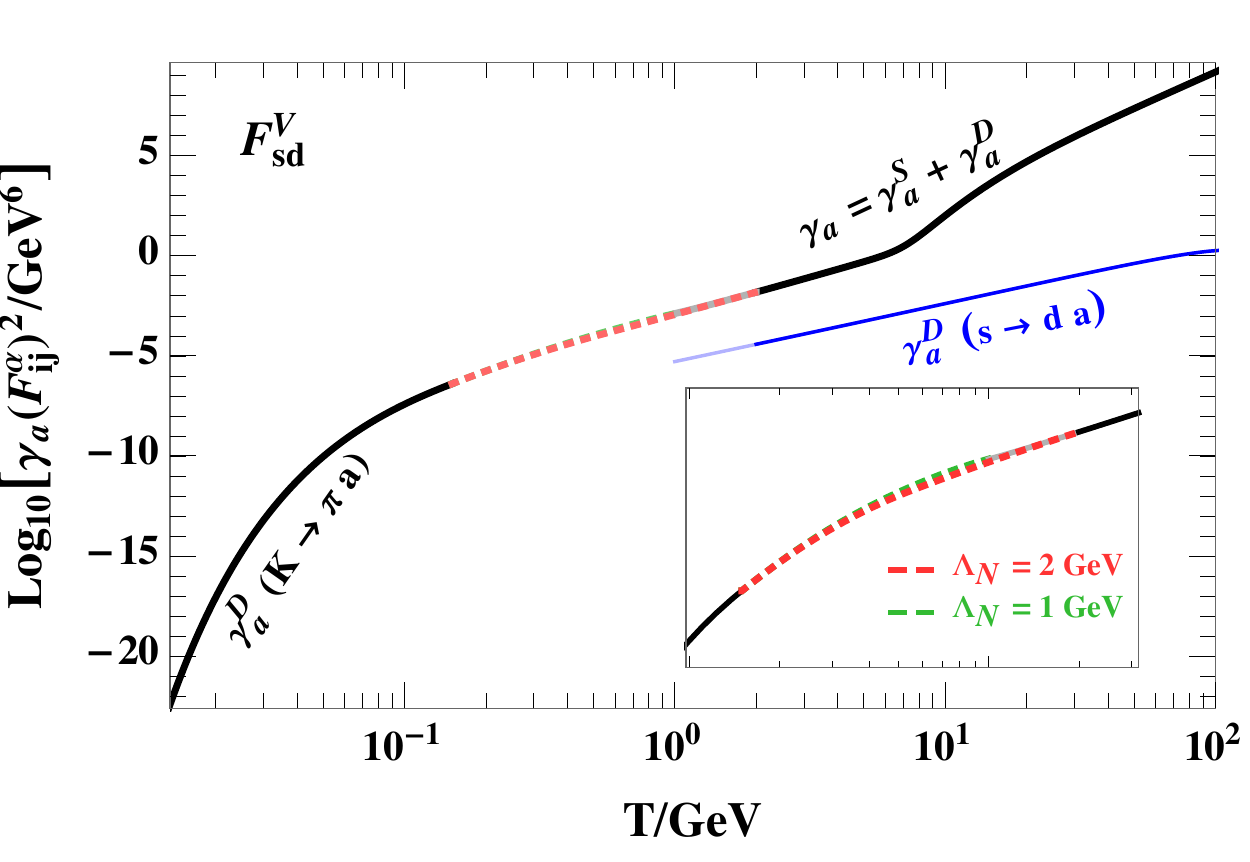}
\includegraphics[width=0.235\textwidth]{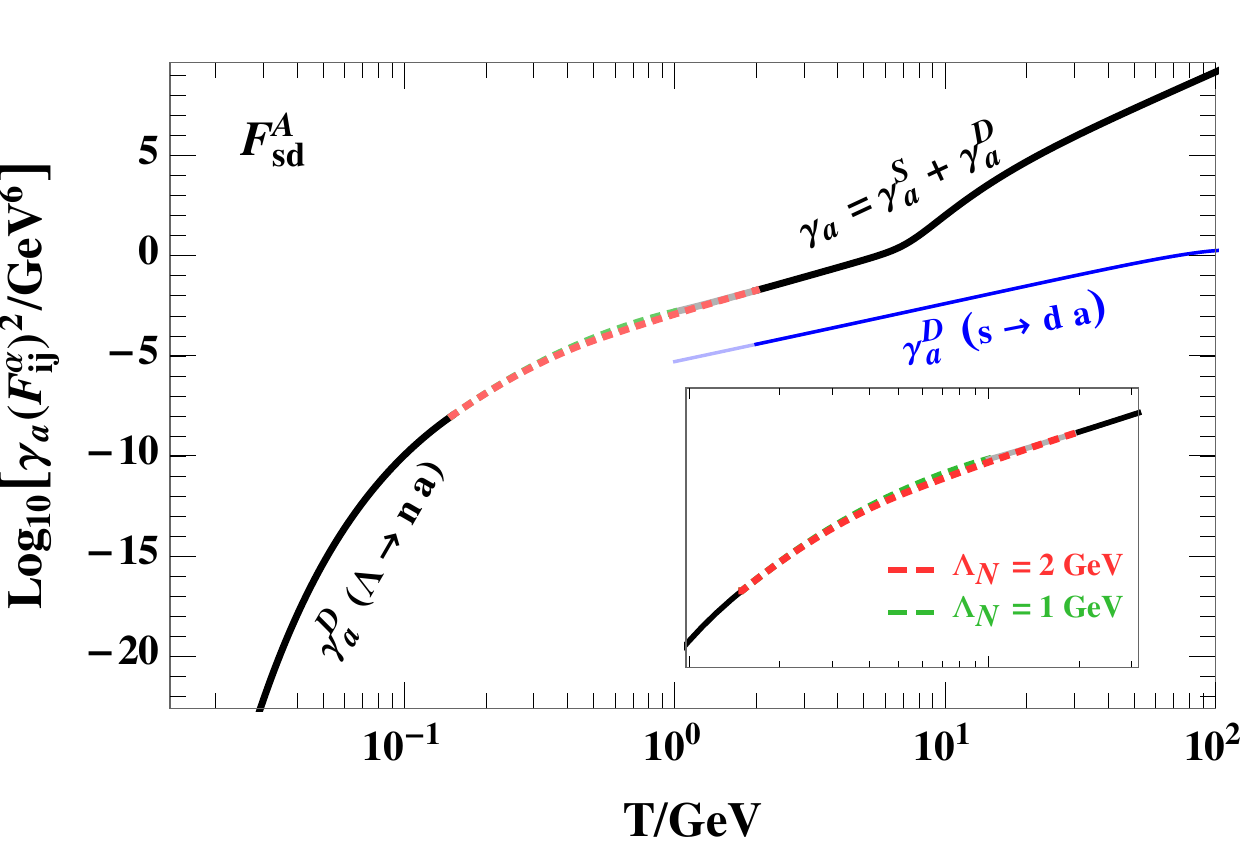}
\includegraphics[width=0.235\textwidth]{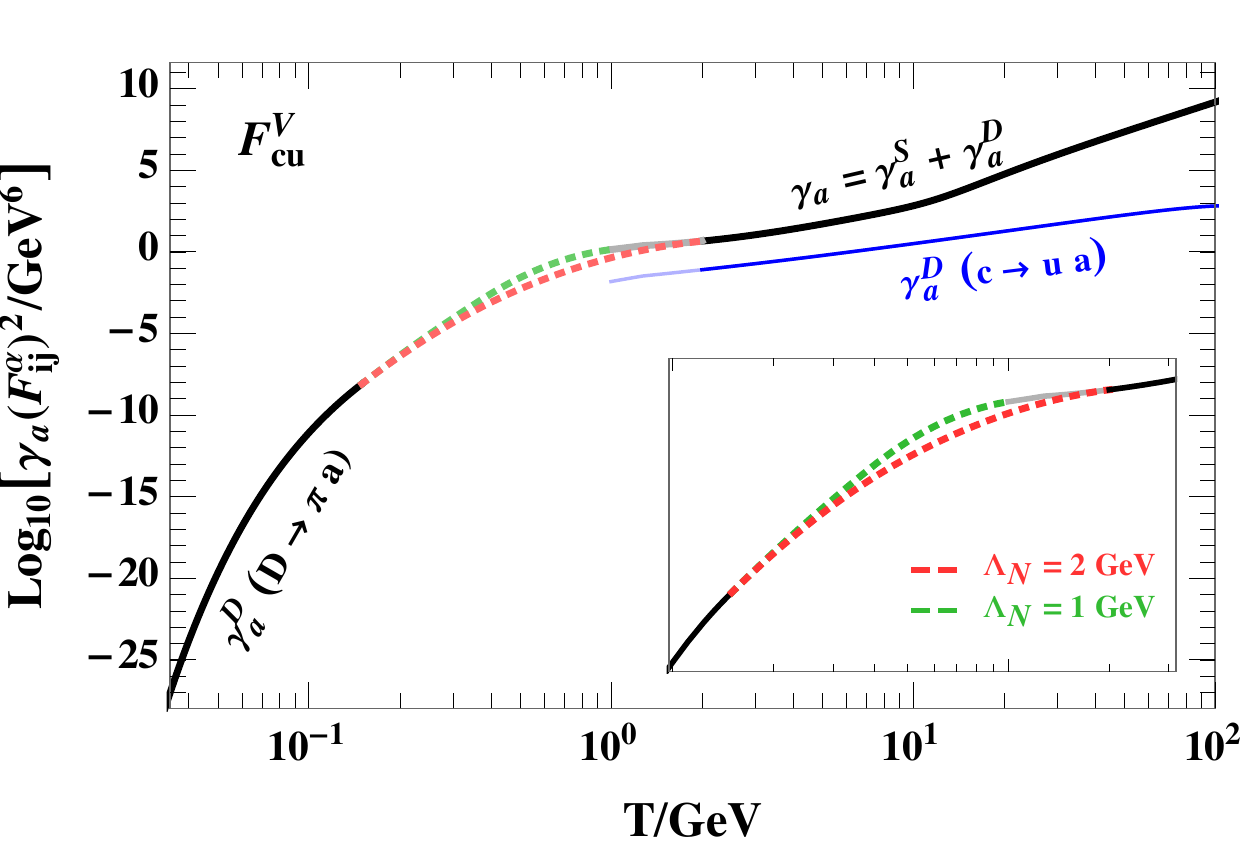}
\includegraphics[width=0.235\textwidth]{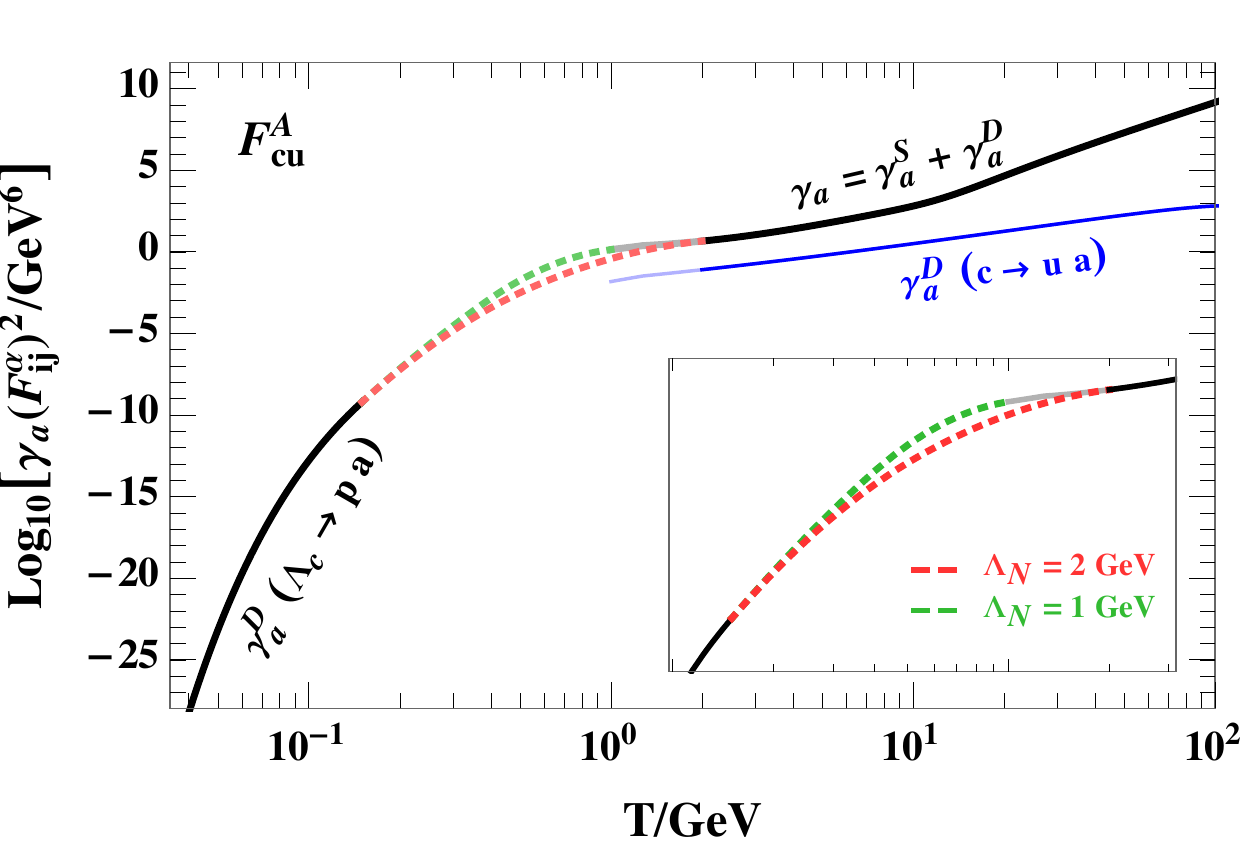}
\caption{\em Axion production rate via FV couplings to second generation quarks. Notation as in Fig.~\ref{fig:Rate3rdQuarks}.}
\label{fig:Rate2rdQuarks}
\end{figure}

In our study, we consider a single FV operator switched on at the time. We evaluate the axion production rate $\gamma_a$, defined as the number of processes per unit time and volume, for couplings to leptons and quarks. The former case, immune from complications as we approach the QCD phase transition (QCDPT), was analyzed by Ref.~\cite{DEramo:2018vss}. The latter was studied by Ref.~\cite{Arias-Aragon:2020shv} only for the third generation quarks since they are heavy enough to work in the deconfined phase. We improved the current state-of-the-art addressing the issues described below.
\begin{itemize}

\item FV interactions mediate axion production via decays and scatterings~\footnote{Expressions for both rates, $\gamma_a = \gamma_a^{\rm D} + \gamma_a^{\rm S}$, can be found in Ref.~\cite{Arias-Aragon:2020shv}. One key difference for scatterings with massless gauge bosons is that FV vector currents are not vanishing since they are proportional to $|m_{\psi_i}-m_{\psi_j}|^2$, leading to a contribution comparable to axial currents.}. The latter contribution is not usually included in this kind of analysis because decays are expected to dominate. In fact, this is never the case. For example, axion production via quark scatterings is proportional to the strong coupling constant $\alpha_s$ that gets large slightly above the QCDPT. We include scatterings properly.

\item For FV couplings to the $b$ quark, we extend the results of Ref.~\cite{Arias-Aragon:2020shv} below the GeV scale. We evaluate the rate from $B$ meson decays below confinement, and we interpolate the results in the intermediate region. Our rates for this case are shown in figure~\ref{fig:Rate3rdQuarks}.

\item We evaluate for the first time the axion production rate due to FV couplings between the first and second quark generations. We perform calculations above and below the confinement scale with perturbative techniques, and we provide a smooth interpolation to connect these two extreme regimes. Our rates for these case are shown in figure~\ref{fig:Rate2rdQuarks}. 

\end{itemize}

\begin{figure}
\centering
\includegraphics[width=0.475\textwidth]{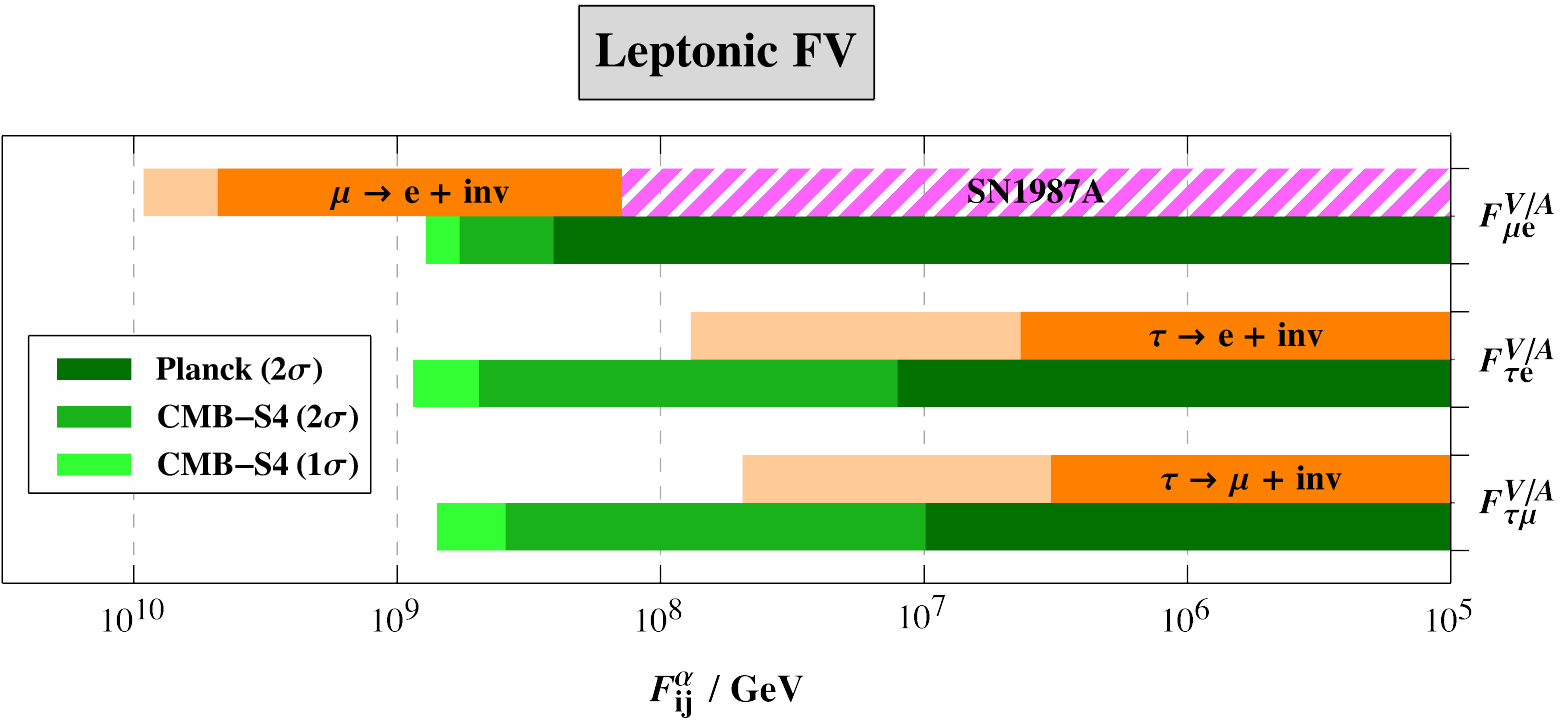} \\
\vspace{0.3cm} 
\includegraphics[width=0.475\textwidth]{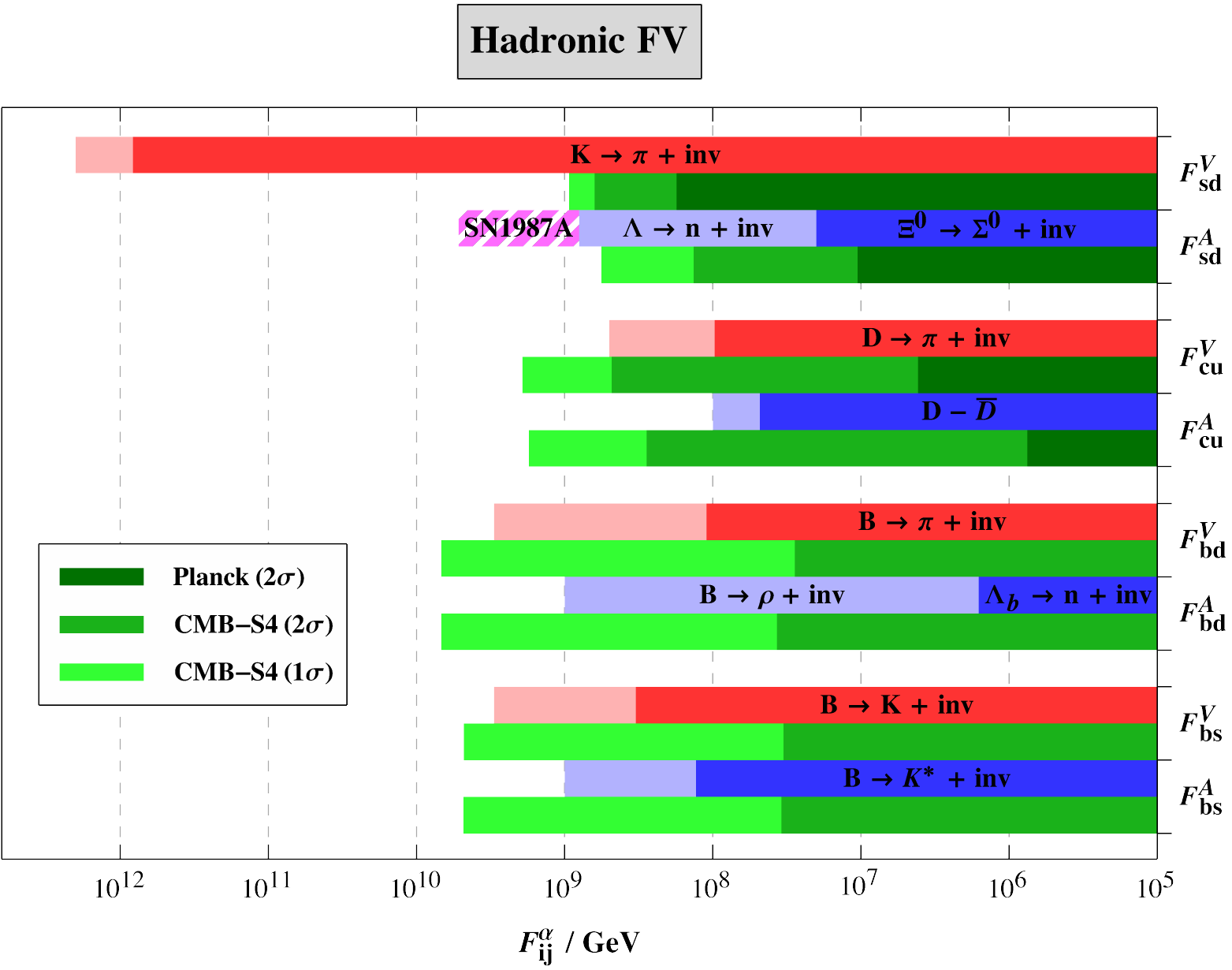}
\caption{\em Terrestrial versus CMB constraints for FV axions coupled to leptons (upper panel) and quarks (lower panel). Darker and fainter areas denote present and future terrestrial bounds, respectively. CMB bounds follow the color code in the legenda, the shaded magenta areas denote SN1987A bounds.}
\label{fig:bounds}
\end{figure}

The rates allow us to evaluate $\Delta N_{\rm eff}$ from different FV couplings. We compare the new cosmological bounds and prospects found in this analysis with current and future constraints from terrestrial experiments. Our cosmological analysis assumes one operator at the time. If multiple interactions are present, as it is often the case for plausible models, there will be several axion production channels and therefore the effective parameters defined in Eq.~\eqref{eq:NormalFV} will have to satisfy more severe constraints. In this respect, our bounds are the most conservative ones. The main findings of our work are summarized in figure~\ref{fig:bounds}.


\vspace{0.2cm}

\noindent {\bf Leptonic Production.} The intricate physics of confinement poses no problem for production via leptons. Charged lepton decays, $\ell^\pm_i \rightarrow \ell^\pm_j a$ with $i \neq j$ and $m_{l_i} > m_{l_j}$, are accounted for in the literature~\cite{Baumann:2016wac,DEramo:2018vss}. However, FV interactions mediate axion production via scatterings of thermal bath particles: lepton/antilepton annihilations ($\ell^\pm_i + \ell^\mp_j \rightarrow \gamma + a$) and Primakoff-like scatterings ($\ell^\pm_{i,j} + \gamma \rightarrow \ell^\pm_{j,i} + a$). The rates are proportional to the electromagnetic fine structure constant $\alpha_{\rm em}$. 

Lepton interactions with the surrounding plasma lead to an effective mass $m_{\ell,{\rm eff}} = (\sqrt{m_\ell^2+4m_{\rm th}^2}+m_\ell )/2$ with the thermal contribution $m_{\rm th}^2 = e^2 T^2/8$~\cite{Petitgirard:1991mf,Chesler:2009yg}. Axion production mediated by the operators in Eq.~\eqref{eq:LaFV} is most efficient at lower temperatures ({\it i.e.}, IR domination), and it is maximized at temperature scales around the heavier lepton mass. A rigorous thermal field theory treatment would not alter the total production rate~\footnote{Things are different for the anomalous axion coupling to gauge bosons where the long range nature of the associated interaction leads to unpleasant IR singularities~\cite{Salvio:2013iaa}}.

\begin{figure}
\centering
\includegraphics[width=0.235\textwidth]{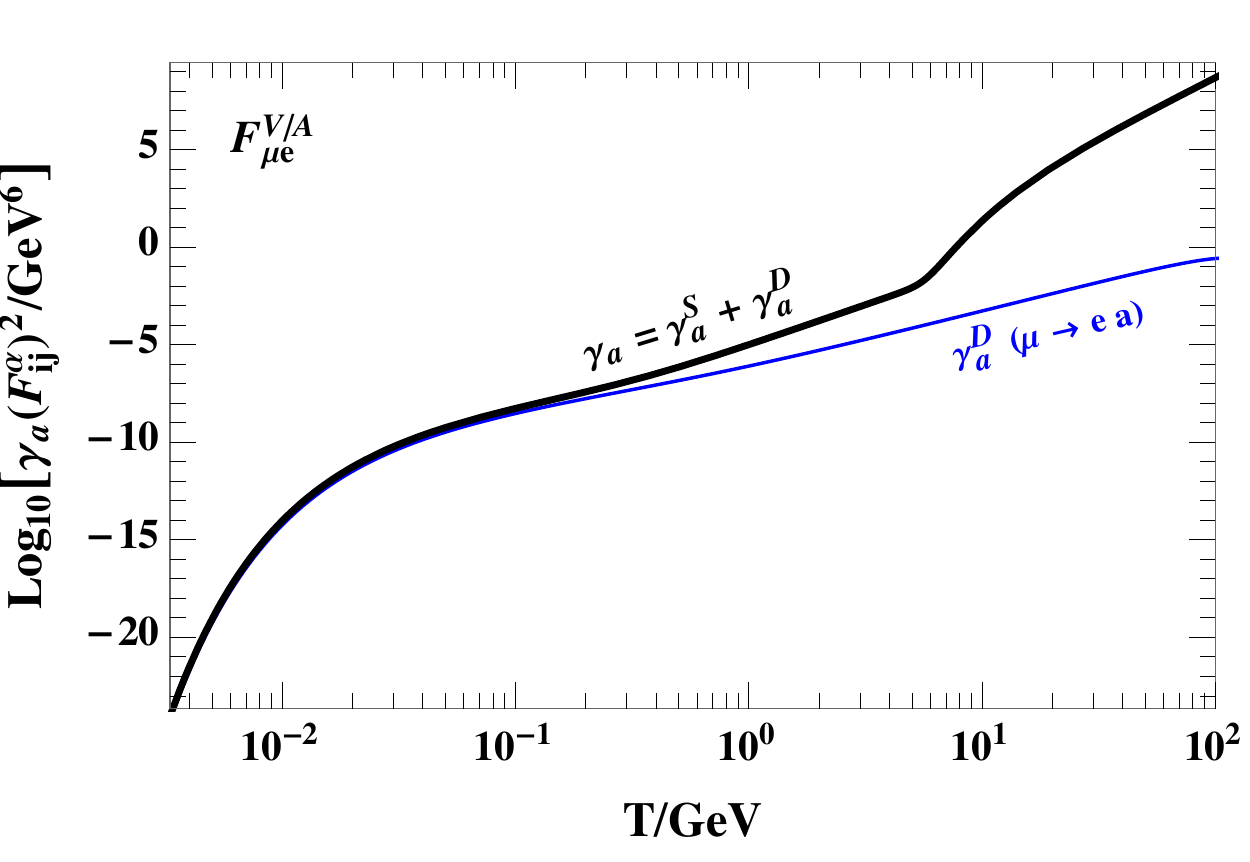}
\includegraphics[width=0.235\textwidth]{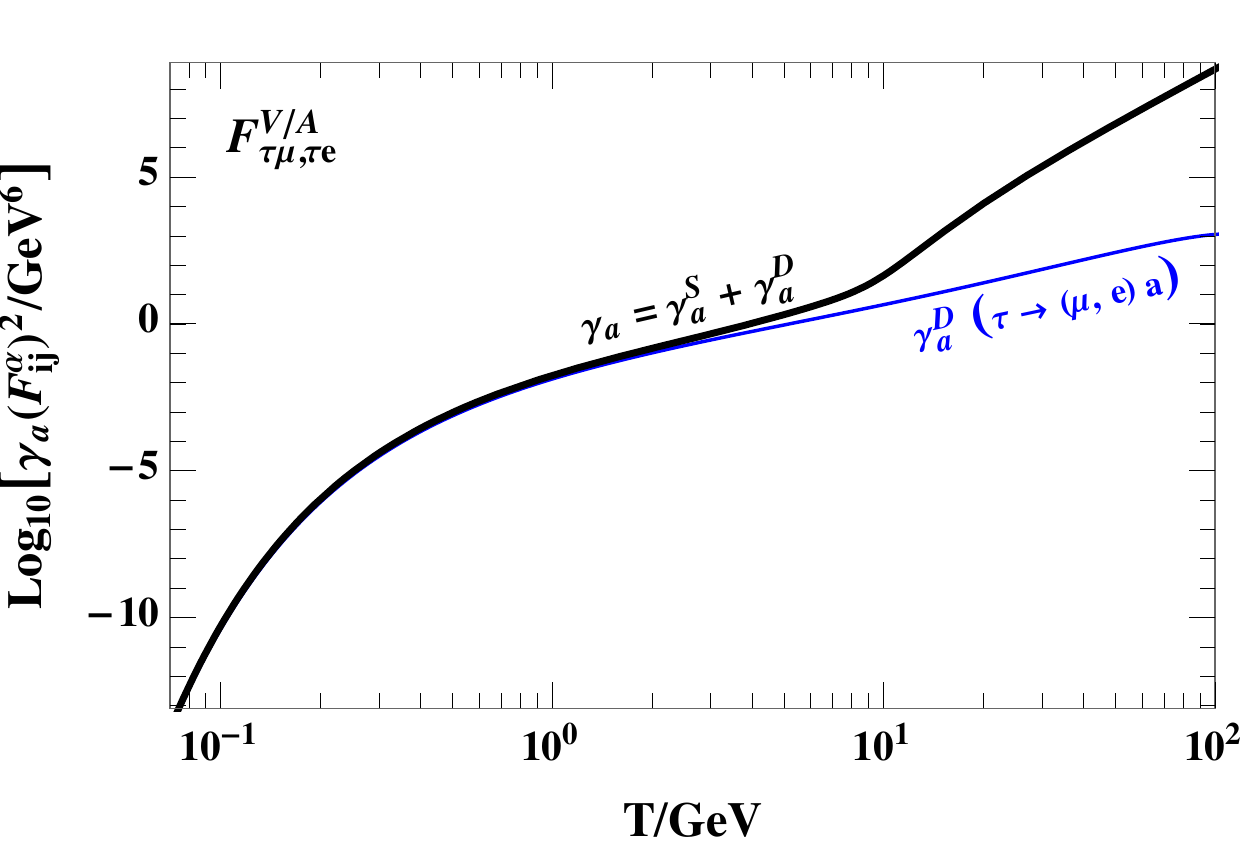}
\caption{\em FV axion production rate from charged leptons for decaying $\mu$ (left) and $\tau$ (right). Black and blue lines denote the total rate and the partial rate via decays, respectively.}
\label{fig:RateSingleLeptonFV}
\end{figure}

The production rates are quantified by the solid black lines in Fig.~\ref{fig:RateSingleLeptonFV} for the cases when the heavier lepton in the interaction is $\mu$ (left panel) or $\tau$ (right panel). For comparison, we illustrate the partial rate contribution from lepton decays with solid blue lines. Once we integrate the Lagrangian in Eq.~\eqref{eq:LaFV} by parts, we find pseudo-Yukawa interactions with strength proportional to $(m_{\ell_j} \mp m_{\ell_i})$ for vector and axial currents, respectively. If we neglect the mass of the lighter lepton, the two couplings give identical rates; the case when corrections to this degeneracy are larger is for $\tau$-$\mu$ couplings where the difference is nearly $30\%$ since $\left(m_\tau + m_\mu\right)^2/\left(m_\tau - m_\mu\right)^2 \approx 1.3$. The results for $\tau$-$\mu$ and $\tau$-$e$ couplings are nearly equivalent for the same reason. We present the results combining these nearly degenerate cases in Fig.~\ref{fig:RateSingleLeptonFV}, but we employ the appropriate rate for each case in our analysis.

At high temperatures, much higher than the mother particle mass, the production rate is controlled by scatterings with gauge bosons. While scatterings with massive weak gauge bosons are relatively suppressed due to their heavy mass, those with massless photons turn out to be dominant contributions to FV axion thermalization below the weak scale. Subsequently, when the thermal bath temperature approaches the mass of the heavier lepton involved in the cubic interaction, decay processes become more and more important to produce axions.


\vspace{0.2cm}

\noindent {\bf Hadronic Production.} The description where the axion is coupled to quarks breaks down once we reach the confinement scale. Computing the axion production rate for FV couplings to quarks is less straightforward.

Third-generation quarks are easier to handle because of their large masses, and this is true, especially, for the top quark. Axion production via FV couplings to the top was studied in Ref.~\cite{Arias-Aragon:2020shv}, and it was found that $\Delta N_{\rm eff}$ is at most barely within reach of future CMB-S4 surveys. The bottom quark is somewhat an intermediate case: production becomes rapidly inefficient once we approach the QCDPT as a consequence of the Maxwell-Boltzmann suppression in the bottom quark number density. This was also studied in Ref.~\cite{Arias-Aragon:2020shv}, but only in the perturbative region and accounting solely for the decay contribution. We improve this former analysis by including scatterings and providing the rate below the confinement scale. 

Perturbative QCD is a reliable computational tool at $T \gtrsim \Lambda_N$, and we employ in our analysis the hadronic matrix elements provided by Ref.~\cite{MartinCamalich:2020dfe} to quantify axion production via hadrons. The Lorentz structure of the FV operator combined with the invariance of strong interactions under parity selects the hadronic final states. The rates are shown in Fig.~\ref{fig:Rate3rdQuarks}, and these numerical results and the ones afterward are obtained with the aid of the `\texttt{RunDec}'~\cite{Chetyrkin:2000yt,Herren:2017osy} code which allows us to run $\alpha_s$ and quark masses up to four loops. Solid black lines indicate the total axion production rate in the two opposite regimes where we have computational control. We consider the two values $\Lambda_N = (1, 2) \,{\rm GeV}$ for the fiducial perturbativity scale. On the opposite side, we account for production via hadron decays up to $T \lesssim \Lambda_{\rm HRG} = 150\,{\rm MeV}$ above which the hadron resonance gas approximation~\cite{Hagedorn:1984hz,Huovinen:2009yb,Megias:2012hk} is not applicable~\cite{Venumadhav:2015pla}. We note that, in the hadronic resonance gas framework, even hadrons with heavy quarks such as charms and bottoms are considered to be thermalized at $\Lambda_{\rm HRG}$. Notwithstanding those hadron masses much heavier than the phase boundary, the so-called statistical hadronization approach is still the valid way to account for their distribution with thermal weights (see Ref.~\cite{Andronic:2017pug} for details). We interpolate in the intermediate region via the cubic spline method, and this is analogous to Refs.~\cite{DEramo:2021psx,DEramo:2021lgb} for the flavor conserving case. The results are shown by dashed lines, and we will comment in the next section how the choice of $\Lambda_N$ does not affect our predictions significantly. The contribution from scatterings overwhelms the one from decays; the former is enhanced by $\alpha_s$, which does not appear in the bottom quark decay width, evaluated at temperatures not too far away from the QCDPT. 

For second generation quarks, confinement plays an even major role. We rely again upon perturbative QCD at $T \gtrsim \Lambda_N$, and we employ the techniques of chiral perturbation theory (ChPT)~\cite{Weinberg:1978kz,Gasser:1983yg,Gasser:1984gg} to describe production in the confined phase. Vector currents mediate pseudo-scalar meson decays to an axion plus a pseudoscalar meson, and axial currents provide vector meson final states. For baryon decays, both currents contribute to the transition amplitudes. The four panels in Fig.~\ref{fig:Rate2rdQuarks} show our numerical results with notation for solid and dashed lines the same as in Fig.~\ref{fig:Rate3rdQuarks}. Even in these cases, the partial production rate from quark decays in the deconfined phase is subdominant with respect to scatterings as a consequence of the $\alpha_s$ enhancement. Thus within the perturbative QCD regime, valid for $T \gtrsim \Lambda_N$, quark and gluon scatterings control axion production. 


\vspace{0.2cm}
 
\noindent {\bf Results.} The rates in Figs.~\ref{fig:RateSingleLeptonFV}-\ref{fig:Rate2rdQuarks} allow us to track the axion number density ($n_a$) via the Boltzmann equation 
\be
\frac{d n_a}{d t} + 3 H n_a = \gamma_a \left( 1 - \frac{n_a}{n_a^{\rm eq}} \right) \ .
\label{eq:BoltzmannEQ}
\ee
The Hubble rate $H$ accounts for the expanding universe, thermal production is accounted for by the right-hand side with $n_a^{\rm eq}$ the equilibrium number density. We run the Boltzmann code with vanishing initial abundance at the weak scale since production is IR dominated. Eventually, the expansion takes over and the comoving number density $Y_a = n_a / s$, where $s$ is the entropy density, reaches a constant value $Y_a^\infty$ that leads to $\Delta N_{\rm eff} \simeq 75.6 \, (Y_a^\infty)^{4/3}$. 

Our main results are summarized in Fig.~\ref{fig:bounds} where we show leptonic and hadronic FV in the upper and lower panels, respectively. The green bars identify cosmological constraints: dark ones denote the current bound from Planck whereas the pale and palest green bars illustrate future prospects at 2$\sigma$ and 1$\sigma$, respectively. All present bounds become more severe as the decaying particles get lighter, and this is because Planck is sensitive to axions that reach thermal equilibrium around the QCDPT; axions that decoupled at a temperature $T_D$ provide a $\Delta N_{\rm eff} \simeq 13.69 \, g_{*s}(T_D)^{-4/3}$ with $g_{*s}$ the number of entropic degrees of freedom~\cite{DEramo:2021lgb}. On the contrary, CMB-S4 prospects feature an opposite behavior because they will be testing axions that do not thermalize, and the abundance is proportional to the square of the interaction strength that scales proportionally to fermion masses. 

The rates in Figs.~\ref{fig:Rate3rdQuarks} and \ref{fig:Rate2rdQuarks} are interpolated across the QCDPT for two values of the fiducial scale $\Lambda_N$. Does this choice impact the  bounds in Fig.~\ref{fig:bounds}? The answer depends on the case. For couplings where the heavier fermion is the strange quark, there is no difference and this is because axion production is efficient well below the scale $\Lambda_N$. The charm quark is an intermediate case since its mass falls right in between the two chosen values for $\Lambda_N$, and we checked that the impact is smaller than a factor of 2 on the cosmological bound. Finally, the bottom quark mass is heavier than both  $\Lambda_N$ and this choice is affecting the bound by less than $50 \%$. The bounds in Fig.~\ref{fig:bounds} are for $\Lambda_N = 2 \, {\rm GeV}$, and this corresponds always to the case where the bound is weaker: we report the most conservative constraints.  Another key ingredient of our analysis is the inclusion of scatterings in the rate, and this has a dramatic effects on the bounds in Fig.~\ref{fig:bounds} for quarks. In particular, our bounds on $F^{\alpha}_{\psi_i \psi_j}$ are approximately one order of magnitude stronger than the one obtained with only decays. The effect is milder for leptons, and including decays improves the bounds only by a factor of two.

FV axions are a plausible origin for the rare lepton and meson decays searched for by terrestrial experiments~\cite{Feng:1997tn,Bjorkeroth:2018dzu,Albrecht:2019zul,MartinCamalich:2020dfe,Calibbi:2020jvd,Bauer:2020jbp,Bauer:2021wjo}. For comparison, we collect in Fig.~\ref{fig:bounds} the current bounds (dark bars) and the future sensitivities (pale bars). We add for completeness the astrophysical constraints on $F_{sd}^A$ and $F_{\mu e}^{V/A}$ with shaded magenta bars. 

The upper panel of Fig.~\ref{fig:bounds} shows the landscape of FV couplings to leptons. Both terrestrial searches and cosmological productions do not discriminate between vector and axial couplings, so we present constraints together and we just distinguish among lepton flavors. Current and future laboratory constraints are visualized by the orange bars. The best bound on $F_{\mu e}^{V/A}$ comes from $\mu \rightarrow e +\,{\rm inv}$~\cite{Jodidio:1986mz}, and the work in Ref.~\cite{Calibbi:2020jvd} suggested that this bound can be improved in the future by the MEGII-fwd experiment~\cite{MEGII:2018kmf}. Alternatively, this can also be searched for by the Mu3e collaboration~\cite{Perrevoort:2018ttp}. We also include the astrophysical SN1987A bound from Ref.~\cite{Calibbi:2020jvd}. Searches for $\tau \rightarrow  \left(e , \mu\right) +\,{\rm inv}$ by the ARGUS collaboration~\cite{ARGUS:1995bjh} constrain both $F_{\tau e}^{V/A}$ and $F_{\tau \mu}^{V/A}$, and Belle II~\cite{RodriguezPerez:2019nhw} will improve the sensitivity on both couplings~\cite{Calibbi:2020jvd}.

The lower panel of Fig.~\ref{fig:bounds} illustrates the current situation and future prospects for FV couplings to quarks. Current bounds for the vector and axial currents are, respectively, represented by the red and blue bars, and the expected sensitivities in forthcoming experiments correspond to the pale colored bars. Searches by the NA62 collaboration~\cite{NA62:2021zjw} for $K^+ \rightarrow \pi^+ + \, {\rm inv}$ provide the best bound on the vector coupling $F_{sd}^V$ with future improvements by KOTO~\cite{MartinCamalich:2020dfe}. The axial coupling $F_{sd}^A$ is instead tested by searches for $\Xi^0 \rightarrow \Sigma^0 + {\rm inv}$~\cite{ParticleDataGroup:2018ovx}, and the BESIII collaboration will provide an improved bound via searches for the hyperon decay $\Lambda \rightarrow n +{\rm inv}$~\cite{Li:2016tlt}. These are milder than the SN1987A constraint~\cite{MartinCamalich:2020dfe,Camalich:2020wac}, the observational cooling signal of which can be reduced by decays of abundant hyperons inside the proto-neutron star~\cite{Oertel:2012qd,Gulminelli:2013qr,Oertel:2016xsn,Fortin:2017dsj}. The recast in Ref.~\cite{MartinCamalich:2020dfe} of the CLEO searches for $D^+ \rightarrow \ell^+ \nu$~\cite{CLEO:2008ffk}, following the strategy proposed by Ref.~\cite{Kamenik:2011vy}, provides the bound on the vector coupling $F_{cu}^V$ through searches for $D^+ \rightarrow \pi^+ +{\rm inv}$, and the BESIII collaboration will improve this constraint by an order of magnitude~\cite{Li:2016tlt}. The axial coupling $F_{cu}^A$ is constrained by $D$-$\bar{D}$ mixing~\cite{HFLAV:2019otj}, and the LHCb Phase II upgrade will improve such a bound~\cite{Kagan:2020vri}. Finally, switching to the third generation, searches by the BaBar collaboration for $B^{+} \rightarrow \pi^{+}+{\rm inv}$ provide the current bound on the vector coupling $F_{bd}^V$~\cite{BaBar:2004xlo}, and Belle II will attain an improved sensitivity to $F_{bd}^V$ by an order of magnitude~\cite{Camalich:2020wac}. A stringent bound on the axial coupling $F_{bd}^A$ arises from the decay channel $\Lambda_b \rightarrow n +{\rm inv}$~\cite{ParticleDataGroup:2018ovx}, and Belle II will improve this bound via $B^{+,0}\rightarrow \rho^{+,0} +{\rm inv}$~\cite{MartinCamalich:2020dfe}. For couplings between second and third generations, we use again the results of the analysis in Ref.~\cite{MartinCamalich:2020dfe}. The BaBar collaboration searches for $B^{+,0}\rightarrow K^{+,0}+{\rm inv}$ and $B^{+,0}\rightarrow K^{* +,0}+{\rm inv}$ give the current experimental bound on $F_{bs}^V$ and $F_{bs}^A$, respectively~\cite{BaBar:2013npw}. Belle II will gather a $100$ larger integrated luminosity compared with BarBar, and this will lead to an order of magnitude enhanced bound on $F_{bs}^{V/A}$.

The comprehensive, yet concise, summary in Fig.~\ref{fig:bounds} shows the complementarity between terrestrial and cosmological searches for axion FV interactions. For couplings to the $\tau$ lepton, cosmological data are already more constraining than the bounds obtained in our laboratories, and this statement will be true in the future as well once the new experiments become operational. Planck bounds cannot compete at the moment with hadronic experiments, and our best constraints on hadronic FV couplings are all due to laboratory searches. Intriguingly, future CMB-S4 will reach a sensitivity stronger than the ones associated with future laboratory experiments. 

The axion is a hypothetical new particle beyond the SM motivated from the top-down, and multiple experimental strategies will probe a large fraction of the coupling parameter space in the near future. CMB data provide an additional and complementary strategy to constrain axion couplings, and they are competitive with other experimental searches for FV couplings. In this study, we restrict ourselves to a model-independent analysis based on the effective operators in Eq.~\eqref{eq:LaFV}. This should be thought of as the low-energy theory valid at the energy scale where axion production takes place, which is typically the mass of the heavier fermion appearing in the interacting vertex. Conceptually, from the high-energy point of view, these FV interactions can arise both for theories where flavor is conserved and FV arises from radiative corrections driven by SM couplings, or for theories where flavor is violated already at the high PQ scale. Our findings summarized by Fig.~\ref{fig:bounds} motivate further studies within UV complete axion models.


\vspace{0.3cm}
\noindent {\it Acknowledgements.} Authors acknowledge E. J. Chun, J. Martin Camalich, L. Merlo, H. Kim, and R. Ziegler for useful discussions. This work is supported by the research grants: ``The Dark Universe: A Synergic Multi-messenger Approach'' number 2017X7X85K under the program PRIN 2017 funded by the Ministero dell'Istruzione, Universit\`a e della Ricerca (MIUR); ``New Theoretical Tools for Axion Cosmology'' under the Supporting TAlent in ReSearch@University of Padova (STARS@UNIPD). The authors also supported by Istituto Nazionale di Fisica Nucleare (INFN) through the Theoretical Astroparticle Physics (TAsP) project. F.D. acknowledges support from the European Union's Horizon 2020 research and innovation programme under the Marie Sk\l odowska-Curie grant agreement No 860881-HIDDeN. The authors acknowledge the Galileo Galilei Institute (GGI) for the kind hospitality during the workshop ``New Physics from the Sky'' while this work was in progress.

\bibliographystyle{apsrev4-2}
\bibliography{HotAxionsFV}

\end{document}